\documentclass[12pt]{article}

\usepackage{epsf}
\usepackage[dvips]{graphicx}
\usepackage{dcolumn}
\usepackage{oldgerm}
\usepackage{amsmath}
\usepackage{psfrag}
\usepackage{color}
\usepackage{latexsym,amsfonts}
\usepackage{graphicx}
\newcommand{\bee}{\begin{eqnarray}}
\newcommand{\eee}{\end{eqnarray}}
\newcommand{\be}{\begin{equation}}
\newcommand{\ee}{\end{equation}}

\begin{document}

\begin{center}
{\bfseries TWO-PHOTON EXCHANGE IN ELECTRON DEUTERON SCATTERING}\\[1.cm]
{\it XX International Baldin Seminar on High Energy Problems\\
{\it RELATIVISTIC NUCLEAR PHYSICS \& QUANTUM CHROMODYNAMICS}\\
Dubna, October 4 ---  9, 2010}

\vskip 5mm

A.P.~Kobushkin$^{1 \dag}$ and Ya.D.~Krivenko-Emetov$^{2}$

\vskip 5mm

{\small
(1) {\it
Bogolyubov Institute for Theoretical Physics\\ Metrologicheskaya Street 14B, 03680 Kiev, Ukraine
}
\\
(2) {
\it Institute for Nuclear Research, Prospekt Nauki 47, 03680 Kiev, Ukraine
}
\\
$\dag$ {\it
E-mail: kobushkin@bitp.kiev.ua}}
\end{center}

\vskip 5mm

\begin{center}
\begin{minipage}{150mm}
\centerline{\bf Abstract}
It is shown that the amplitude of elastic $ed$ scattering beyond Born approximation contains six generalized form factors, but only three linearly independent combinations of them (generalized charge, quadrupole and magnetic form factors) contribute to the reaction cross section in the second order perturbation theory. We examine two-photon exchange and find that it includes two types of diagrams, when two virtual photons interact with the same nucleon and when the photons interact with different nucleons.We discuss contribution of the two-photon exchange in reaction observables, generalized $\mathcal A$ and $\mathcal B$ structure functions and tensor polarization of the deuteron.
\end{minipage}
\end{center}
\vskip 10mm

\section{\label{sec:Introduction}Introduction}
A study of electromagnetic structure of the deuteron, the simplest nucleon system, provides with important information about nucleon-nucleon interaction. Because the deuteron is a spin-1 system its electromagnetic current characterized by three form factors, charge $G_C$, quadrupole $G_Q$ and magnetic $G_M$ form factors. Due to smallness of the fine structure constant $\alpha$ the form factors are usually extracted from experimentally measurable observables in the framework of Born approximation (one-photon exchange, OPE). Nevertheless theoretical calculations \cite{Lev,GakhTomasi,DongChen,KK-ED} show that effects beyond OPE may significantly change results of such procedure.

In what follows we calculate amplitude of two-photon exchange (TPE) of the elastic $ed$ scattering, one of the mostly important effects beyond OPE, and estimate TPE contribution in observables of the process. 
\section{\label{sec:1}Observables beyond one-photon exchange}
From $P$ and $T$ invariance it follows that elastic scattering amplitude of a spin-$\frac12$ particle (electron) on a spin-1 particle (deuteron) is determined by 12 invariant amplitudes. Putting the electron mass to zero reduces the number of the invariant amplitude (form factors) to 6 and the spin structure of the amplitude in the Breit frame may be specified by the following parametrization:
\be\label{Gen.Reduce_Amplitue}
T_{\lambda'\lambda; h}=\left(
\begin{array}{ccc}
\mathcal G_{11}\cos\frac{\theta}2&-\sqrt\frac{\eta}{2}\mathcal G_{10}^{h} & \mathcal G_{1,-1}^{h}\\
\sqrt\frac{\eta}{2}\mathcal G_{10}^{-h} &\mathcal G_{00}\cos\frac{\theta}2&-\sqrt\frac{\eta}{2}\mathcal G_{10}^{h}\\
\mathcal G_{1,-1}^{-h} & \sqrt\frac{\eta}{2}\mathcal G_{10}^{-h}& \mathcal G_{11}\cos\frac{\theta}2
\end{array}
\right).
\ee
Here $T_{\lambda'\lambda; h}$ is reduced amplitude, which is connected with the usual amplitude by
\be \label{Gen.Amplitue}
\mathcal M=\frac{16\pi\alpha}{Q^2}E_eE_d T_{\lambda'\lambda,h},
\ee
$\lambda$ and $\lambda'$ are spin projections of the deuteron and $h$ is sign of electron helicity; $E_e$ and $E_d$ are electron and deuteron energies and $\theta$ is the scattering angle in the Breit system; $\eta=Q^2/4m_d^2$; 
\be \label{Gen.Gmn}
\mathcal G_{10}^{h}=f_1+h\sin\tfrac{\theta}2 f_2, \qquad
\mathcal G_{1,-1}^{h}=f_3+h\sin\tfrac{\theta}2 f_4.
\ee
The form factors $\mathcal G_{11}$, $\mathcal G_{00}$, $f_1$, ..., $f_4$ are complex functions of the two independent kinematical variables, for example $Q^2$ and $\theta$.

In Ref.~\cite{KK-ED} instead of the form factors  $\mathcal G_{11}$, $\mathcal G_{00}$, $f_1$, ..., $f_4$ the following their linear combinations were introduced
\be \label{GcGqGm}
\begin{split}
&\mathcal G_{11}=\mathcal G_C-\tfrac23\eta \mathcal G_Q,\qquad
\mathcal G_{00}=\mathcal G_C+\tfrac43\eta \mathcal G_Q,\\
&f_1=\mathcal G_M + g_1\sin^2\tfrac{\theta}2,\qquad f_2=\mathcal G_M - g_1,\\
&f_3=g_2,\qquad f_4=g_3.
\end{split}
\ee
We call $\mathcal G_Q(Q^2,\theta)$ and $\mathcal G_M(Q^2,\theta)$ the generalized electric, quadrupole and magnetic  from factors.

By standard calculations one derives the differential cross section and components of tensor polarization of the deuteron
\be\label{Cross-s}
\begin{split}
\dfrac{d\sigma}{d\Omega}=&\sigma_{\mathrm M}\mathcal S,\\
%
%
t_{20}=&\dfrac{-\eta}{3\sqrt{2}}\frac{[8(\Re \mathrm e\mathcal G_C^\ast \mathcal G_Q+\frac{\eta}{3}|\mathcal G_Q|^2)+|\mathcal G_M|^2(1+2\mathrm{tg}^2\tfrac{\theta}2) ]}{\mathcal S},\\
t_{21}=&\frac{\sqrt{\eta }}{\sqrt{3}\mathcal S}  \left[
   -2 \cos\tfrac{\theta }{2}\eta \Re \mathrm e \mathcal G_M^\ast  \mathcal G_Q
  -G_M\Re \mathrm e\left(\sin ^2\tfrac{\theta}{2} g_3 + g_2\right)
   -2 \sin ^2\tfrac{\theta }{2}\cos\tfrac{\theta }{2}\eta \Re\mathrm e g_1 G_Q \right],\\
t_{22}=&\frac{-\cos^2\tfrac{\theta }{2} \eta  |\mathcal G_M|^2 
   -4 \sin ^2\tfrac{\theta }{2}\eta G_M \Re \mathrm e g_1 
   +4\cos \tfrac{\theta }{2} (G_C -\tfrac23\eta G_Q)\Re\mathrm e g_2}
   {2\sqrt{3}\cos^2\tfrac{\theta }{2}\mathcal S}.
\end{split}
\ee
In Eqs. (\ref{Cross-s}) 
$\sigma_{\mathrm M}$ is the Mott cross section,
\be\label{S}
\begin{split}
&\mathcal S=\mathcal A+\mathcal B\mathrm{tg}^2\left( \tfrac12\theta_{\mathrm{LAB}}\right),\\
&\mathcal A(Q^2,\theta)=|\mathcal G_C(Q^2,\theta)|^2+\tfrac89\eta^2|\mathcal G_Q(Q^2,\theta)|^2 +\tfrac23\eta|\mathcal G_M(Q^2,\theta)|^2, 
\\
&\mathcal B(Q^2,\theta)=\tfrac43 (1+\eta)\eta|\mathcal G_M(Q^2,\theta)|^2
\end{split}
\ee
and 
$\mathcal G_K^\ast \mathcal G_L=G_K G_L + \delta\mathcal G_K^\ast \mathcal G_L+ 
\mathcal G_K^\ast \delta\mathcal G_L$, $(K,L)=C,Q,M$.

The advantage of using the form factors $\mathcal G_C$, $\mathcal G_Q$ and $\mathcal G_M$ is that the 
expression for the cross section and $t_{20}$ have the same form as in OPE approximation. Nevertheless the Rosenbluth separation of the structure functions $\mathcal A(Q^2,\theta)$ and $\mathcal B(Q^2,\theta)$ can no longer be done because they depend on two variables.

\begin{figure}[t]
 \epsfysize=85mm
 \centerline{
 \epsfbox{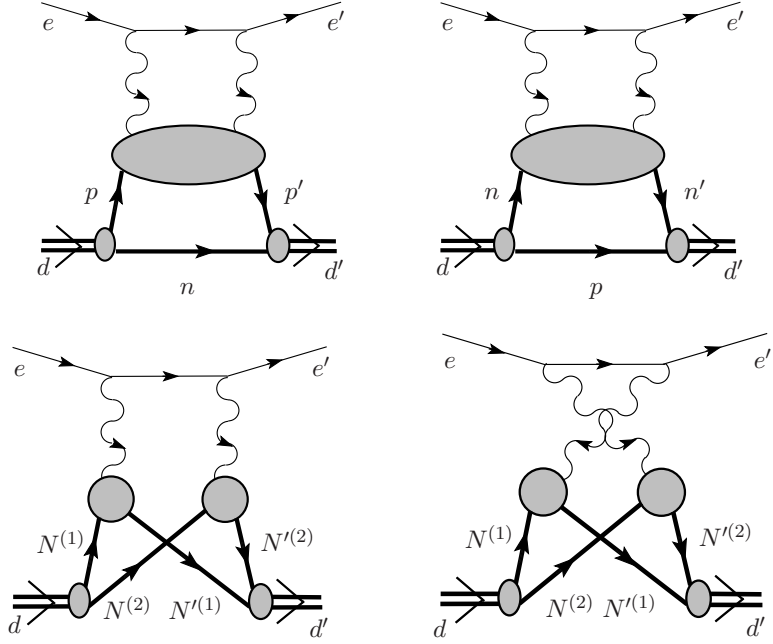}}
 \caption{Two-photon exchange diagrams. The top diagrams correspond to the amplitudes $\mathcal M_p^{\mathrm I}$ and $\mathcal M_n^{\mathrm I}$, the bottom diagrams to the amplitudes $\mathcal M_{\mathrm P}^{\mathrm{II}}$ and $\mathcal M_{\mathrm X}^{\mathrm{II}}$.}
 \label{fig:2photon}
\end{figure}
\section{Calculation of the two-photon exchange\label{two-photon}}
In our calculation of TPE we consider two types of diagrams, where the virtual photons interact directly with the nucleons
\be \label{TPE-definition}
\mathcal M_2=\mathcal M^{\mathrm{I}}+\mathcal M^{\mathrm{II}}.
\ee
One of them, $\mathcal M^{\mathrm{I}}=\mathcal M^{\mathrm{I}}_p+\mathcal M^{\mathrm{I}}_n$, corresponds to diagrams, where both photons interact with the same nucleon (Fig.~\ref{fig:2photon}, top). The other type, $\mathcal M^{\mathrm{II}}=\mathcal M^{\mathrm{II}}_{\mathrm P}+\mathcal M^{\mathrm{II}}_{\mathrm X}$, corresponds to the diagrams, where the photons interact with different nucleons (Fig.~\ref{fig:2photon}, bottom).

Important input in calculation of $\mathcal M^{\mathrm{I}}$ is TPE amplitude for a nucleon $N$. It has the following structure \cite{Guichon}
\be\label{TPE_N}
\mathcal{M}_{2\gamma N}=\frac{4\pi\alpha}{Q^2}\bar u'_h\gamma_\mu u_h \left\langle \vec p{\,'}_{\!\!\!N}\sigma'\left|\widehat H^\mu_N \right|\vec p_N\,\sigma\right\rangle ,
\ee
where $\widehat H^\mu_N$ is the ``effective hadron current''
\be
\widehat H^\mu_N=\Delta\widetilde F_{1}^N\gamma^\mu -\Delta\widetilde F_{2}^N[\gamma^\mu,\gamma^\nu]\frac{q_\nu}{4m_N}+
\widetilde F_{3}^NK_\nu\gamma^\nu\frac{P^\mu}{m_N^2}.
\label{TPE.I.3}
\ee
In Eqs.~(\ref{TPE_N}) and (\ref{TPE.I.3}) $p_N$ and $p{\,'}_{\!\!\!N}$ are the nucleon momenta, $\sigma$ and $\sigma'$ are the nucleon spin projections, $\left|\vec p_N\,\sigma\right\rangle $ and $\left|\vec p{\,'}_{\!\!\!N}\,\sigma'\right\rangle $ are the nucleon spinors, $K=(k+k')/2$, $P=(p_N+p{\,'}_{\!\!\!N})/2$; $\Delta\widetilde F_{1}^N$ and $\Delta\widetilde F_{2}^N$ may be called corrections to the Dirac and Pauli form factors of nucleon $N$ and $\widetilde F_{3}^N$ is a new form factor. All the quantities $\Delta\widetilde F_{1}^N$, $\Delta\widetilde F_{2}^N$ and $\widetilde F_{3}^N$ are of order $\alpha$. They are complex functions of two kinematical variables, e.g. $Q^2$ and $\nu=4PK$.

Considering the deuteron structure nonrelativistically one gets the following expressions for appropriate TPE form factors  for the elastic $ed$ scattering \cite{KK-ED}
\bee
&&\delta \mathcal G_C^{\mathrm I} = 2\delta\mathcal G_E^S\left[I_{00}^0(Q)+I_{22}^0(Q)\right],\quad
\delta \mathcal G_Q^{\mathrm I} = \frac{3\sqrt 2}{\eta}\delta\mathcal G_E^S\left[I_{20}^2(Q)-\frac1{2\sqrt 2}I_{22}^2(Q)\right],\nonumber\\
&&\delta \mathcal G_M^{\mathrm I} = \frac{M}m\left\{\tfrac32\delta\mathcal G_E^S\left[I_{22}^0(Q)+I_{22}^2(Q)\right]+
2\delta\mathcal G_M^S\left[I_{00}^0(Q)-\tfrac12 I_{22}^0(Q)+\sqrt{\tfrac12} I_{20}^2(Q)+\tfrac12 I_{22}^2(Q)\right]\right\},\nonumber\\
&&
g_1^\mathrm{I}=-\epsilon\frac{E_e}{m}\mathcal F_3,\quad g_2^\mathrm{I}=g_3^\mathrm{I}=0,\nonumber
\eee
where
$
\mathcal F_3 =2\frac{M}{m}\widetilde F_3^S\left[I_{00}^0(Q)-\tfrac12 I_{22}^0(Q)+\sqrt{\tfrac12} I_{20}^2(Q)+\tfrac12 I_{22}^2(Q)\right]$, generalized nucleon electric and magnetic form factors are defined by (see Ref~\cite{BorisyukKob_phen})
\be
\delta \mathcal G_E^N= \Delta\widetilde F_1^N -\tau \Delta\widetilde F_2^N+\frac{\nu}{4m_N^2}\widetilde F_3^N,\quad
\delta \mathcal G_M^N=\Delta\widetilde F_1^N + \Delta\widetilde F_2^N+\frac{\epsilon\nu}{4m_N^2}\widetilde F_3^N,\nonumber
\ee
$\tau \approx 4\eta$, $\nu\approx m_N E_e$ and $\epsilon
$ is the commonly used polarization parameter, $I_{\ell'\ell}^{L}(Q)=\int_0^\infty dr j_L\left(\tfrac{1}{2}Qr\right)u_{\ell'}(r)u_{\ell}(r)$, $u_\ell(r)$ is the radial deuteron wave function for orbital momentum $\ell$ and $\delta\mathcal G_E^S=\frac12(\delta\mathcal G_E^p + \delta\mathcal G_E^n)$, etc.

\begin{figure}[t]
 \epsfysize=95mm
 \centerline{\epsfbox{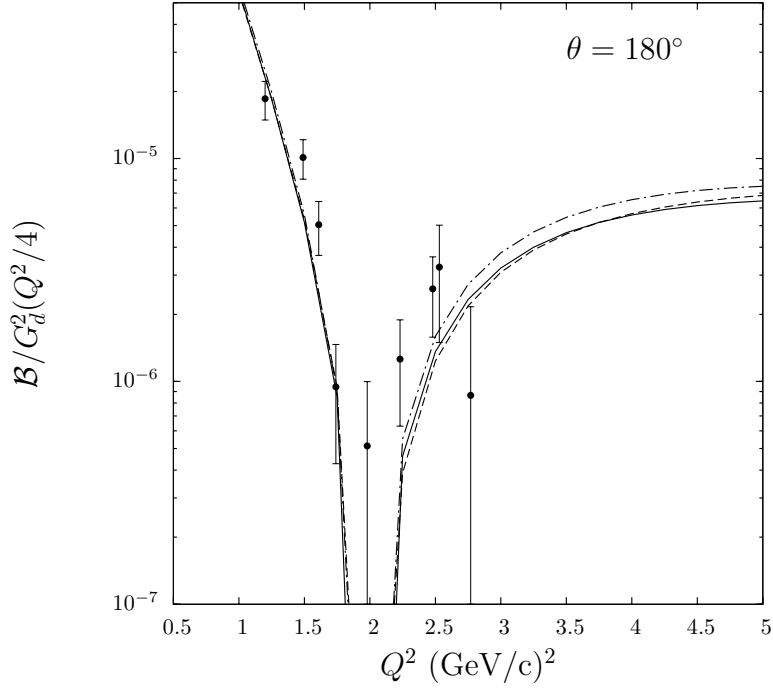}}
 \caption{Results of calculations for $\mathcal B$ at $\theta_\mathrm{LAB}=180^\circ$ (the curves are explained  in the text). $G_d(t)=(1+t/0.71)^{-2}$ is dipole form factor. Experimental data are from \cite{Bosted}.}
 \label{fig:180}
\end{figure}

The amplitude $\mathcal M^{\mathrm{II}}$ was calculated within hard-photon approximation, i.e. assuming that each intermediate photon carries about half of the transferred momentum $\Delta_1\sim \Delta_2\sim \frac{q}2$. Omitting tedious calculations (see Ref.~\cite{KK-ED}), we obtain
\be
\label{FF_II}
\begin{split}
&\delta\mathcal G_C^\mathrm{II}=\kappa\left(G_{EE} -\tfrac13\eta G_{MM}\right),\quad
\delta\mathcal G_Q^\mathrm{II}=-\dfrac{\kappa}{2}G_{MM},\\
&\delta\mathcal G_M^\mathrm{II}=\dfrac{2\kappa G_{EM}}{1+\sin^2\tfrac{\theta}2},\quad
g_1^\mathrm{II}=\dfrac{\kappa G_{EM}\cos^2\tfrac{\theta}2}{1+\sin^2\tfrac{\theta}2},\quad
g_2^\mathrm{II}=g_3^\mathrm{II}=\kappa\eta\cos\tfrac{\theta}2 G_{MM}.
\end{split}
\ee
Here 
\be
\begin{split}
&G_{EE}=G_E^p(\tfrac14Q^2)G_E^n(\tfrac14Q^2),\quad
G_{MM}=G_M^p(\tfrac14Q^2)G_M^n(\tfrac14Q^2),\\
&G_{EM}=\tfrac12\left[G_E^p(\tfrac14Q^2)G_M^n(\tfrac14Q^2) +G_M^p(\tfrac14Q^2)G_E^n(\tfrac14Q^2)\right]
\end{split}
\ee
and 
\be\label{kappa}
\kappa=-\dfrac{128\alpha E_e}{Q^4}\,\mathcal C,\quad \text{where}\quad 
\mathcal C=\frac{1}{(2\pi)^3}\int\frac{d^3pd^3p\,' U_0(p)U_0( p')}
{1+\dfrac{4E_e}{Qm_d}(p_z+p'_z)-8\cos\tfrac{\theta}2\dfrac{E_e(p_x-p\,'_x)}{Q^2} +i0}.
\ee
In Eq.~(\ref{kappa}) $U_0(p)$ is $S$-component of the deuteron wave function in the momentum representation. To evaluate the integral above one can use the integral representation for the propagator
$\dfrac{1}{\alpha +i0}=-i\int_0^\infty d\tau e^{i(\alpha +i0)\tau}$ and reduce $\mathcal C$ to a one-dimensional integral
\be\label{S_prime_1}
\mathcal C=-if\int_0^\infty \dfrac{dy}{y^2}e^{ify} u^2_0(y),\quad \text{where}\quad
f=Q^2\left[ 4E_e\sqrt{4\cos^2\tfrac{\theta}2 +\frac{Q^2}{m_d^2}}\right]^{-1} .
\ee
\section{\label{numerical}Numerical calculations and conclusions}
\begin{figure}[t]
 \epsfysize=95mm
\centerline{\epsfbox{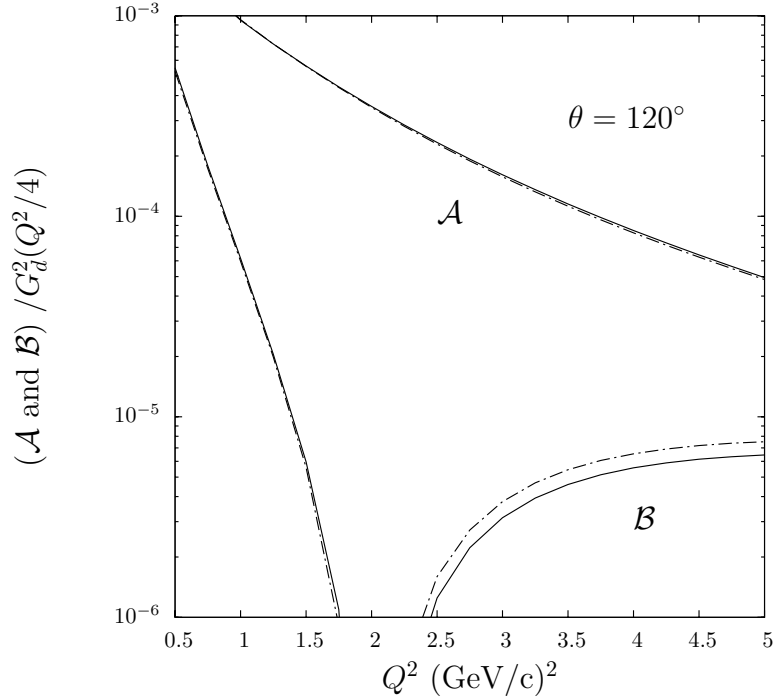}}
 \caption{$\mathcal A$ and $\mathcal B$ at $\theta_\mathrm{LAB}=120^\circ$. Full curve for ONE+TPE (TPE is calculated with CD-Bonn deuteron wave function), dot-dashed for ONE approximation.}
 \label{fig:AB}
\end{figure}
\begin{figure}[t]
 \epsfysize=85mm
\centerline{\epsfbox{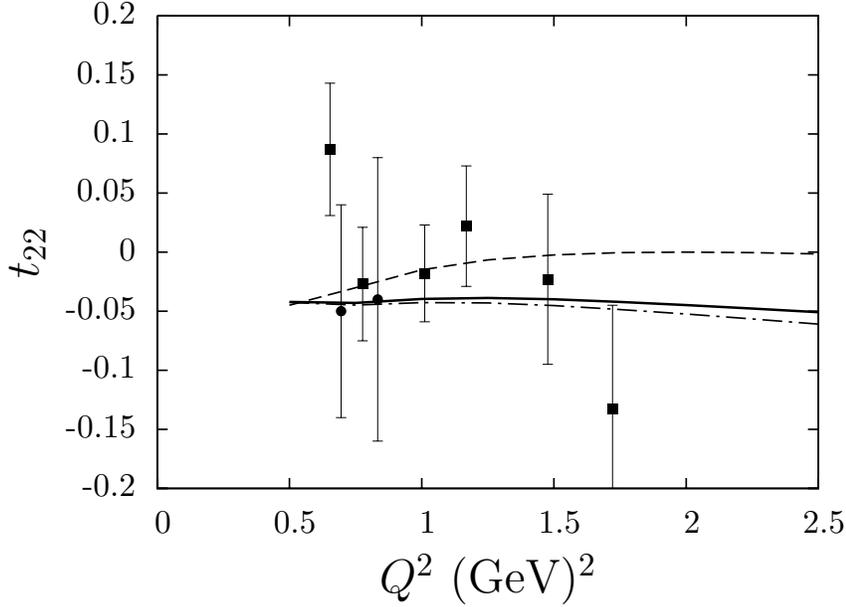}}
 \caption{$t_{22}$ at $\theta_\mathrm{LAB}=70^\circ$. Dashed line is for OPE-approximation, full and dashed-dot lines are for OPE+TPE with TPE calculated with CD-Bonn and Paris deuteron wave functions, respectively. Data are from \cite{Abbot} and \cite{Garcon} (circles and boxes, respectively).}
 \label{fig:t21}
\end{figure}
In Figure~\ref{fig:180} we compare results of our calculations for $\mathcal B$ at $\theta_\mathrm{LAB}=180^\circ$ with ONE analysis of Ref.~\cite{KobushkinSyamptomov} (dot-dashed). For TPE calculations we have used the deuteron wave functions for CD-Bonn and Paris potentials.
ONE+TPE with TPE calculated with CD-Bonn deuteron \cite{CD-Bonn} wave function and Paris \cite{Paris} deuteron wave function are given by full and dashed curves, respectively. One sees that at $Q^2>2\ \mathrm{GeV}^2$ TPE contribution becomes more than 10\% in $\mathcal B$, while in $\mathcal A$ it is not significant, see Figure~\ref{fig:AB}.

TPE effect in $t_{22}$ was found significant (Figure~\ref{fig:t21}), but in $t_{20}$ and $t_{21}$ its contribution is minor ($<$1\%).

%

\end{document}